\documentclass[12pt]{article}
\usepackage{epsfig,amsmath,amsfonts,amssymb,makeidx,ifthen}
\makeatletter
\@addtoreset{equation}{section}
\makeatother

\makeatletter
\long\def\@makecaption#1#2{{\small
\advance\leftskip1cm
\advance\rightskip1cm
\vskip\abovecaptionskip
\sbox\@tempboxa{#1: #2}%
\ifdim \wd\@tempboxa >\hsize
 #1: #2\par
\else
\global \@minipagefalse
\hb@xt@\hsize{\hfil\box\@tempboxa\hfil}%
\fi
\vskip\belowcaptionskip}}
\makeatother
\def\eq#1\en{\begin{equation}#1\end{equation}}  
\def\eqa#1\ena{\begin{align}#1\end{align}}
\def\eqg#1\eng{\begin{gather}#1\end{gather}}
\newcommand{\lb}[1]{\label{e:#1}}
\newcommand{\rlb}[1]{\eqref{e:#1}} 
\newcommand{\nl}{\notag\\}

\newcommand{\sbkt}[1]{\langle#1\rangle}

\newcommand{\Bbkt}[1]{\Bigl\langle#1\Bigr\rangle}

\newcommand{\meas}{\Pi}
\newcommand{\rhocan}{\rho_\mathrm{can}}
\newcommand{\Tr}{\operatorname{Tr}}
\newcommand{\tp}{\tilde{p}}
\newcommand{\tgamma}{\tilde{\gamma}}
\newcommand{\err}{\epsilon}

\begin{document}
\noindent
{\bf\large Quantum Jarzynski-Sagawa-Ueda relations}
\par\bigskip

\noindent
{\small Yohei Morikuni${}^1$ and Hal Tasaki\footnote{
Department of Physics, Gakushuin University, Mejiro, Toshima-ku, Tokyo 171-8588,
 Japan.
}}

\begin{abstract}
We consider a (small) quantum mechanical system which is operated by an external agent, who changes the Hamiltonian of the system according to a fixed scenario.
In particular we assume that the agent (who may be called a demon) performs measurement followed by feedback, i.e., it makes a measurement of the system and changes the protocol according to the outcome.
We extend to this  setting the generalized Jarzynski relations, recently derived by Sagawa and Ueda for classical systems with feedback.
One of the two relations by Sagawa and Ueda is derived here in error-free quantum processes, while the other is derived only when the measurement process involves classical errors.
The first relation leads to a second law which takes into account the efficiency of the feedback.
\end{abstract}

\section{Introduction}
Recent progress in statistical physics has led to nontrivial exact relations such as the fluctuation theorem \cite{ECM,Giovanni,Crooks} and the Jarzynski relation \cite{Chris,Crooks}, which hold even when physical systems are driven out of equilibrium.
These relations reveal rich structures hidden in the canonical formalism of equilibrium states, and also suggest that there can be some universal structures out of equilibrium.
Since these exact relations mainly deal with fluctuation, they are expected to find useful applications in small systems, where relevant energy scale is comparable to that of thermal fluctuation.

In recent papers \cite{SU1,SU}, Sagawa and Ueda studied the effect of feedback or the Maxwell demon in (small) thermodynamic systems.
They showed that the second law of thermodynamics and the Jarzynski relation should be properly modified in order to take into account the information of the system gained by the demon.
In a recent beautiful experiment, Toyabe, Sagawa, Ueda, Muneyuki, and Sano \cite{Toyabe} have designed a system with an artificial demon intervening the time evolution, and  clearly demonstrated that one can convert information into useful work\footnote{
The term ``information-to-energy conversion'' found in the title of \cite{Toyabe} should better be read ``information-to-free-energy conversion.''
}.
They also confirmed the validity of one of the generalized Jarzynski relations obtained by Sagawa and Ueda \cite{SU}, which we shall call Jarzynski-Sagawa-Ueda relations.

In the present paper, we treat a quantum system with feedback, and look for extensions of the Jarzynski-Sagawa-Ueda relations, which were proved only for classical systems.
There may be a possibility that such extensions will become relevant to experiments or manipulations of small quantum systems.
But let us stress that by studying quantum extensions one will get a better understanding of basic structures of the relations, e.g.,  which relation is universal and which is intrinsic to classical systems.
We believe that such an attempt is of interest from a purely theoretical point of view.
In fact we found that one of the two relations derived by Sagawa and Ueda can be derived in an error-free quantum process, while the other can be (for the moment) shown only when we include somewhat artificial classical errors into the process.
For other related works on processes including feedback and/or measurement, see \cite{CTH,JS} and references therein.

\bigskip
The present paper is organized as follows.
In section~\ref{s:noerror}, we study the most basic setting with no errors and derive the quantum extension of one of the two relations by Sagawa and Ueda.
In section~\ref{s:error}, we introduce classical errors to the feedback process, and derive a quantum counterpart of the other relation.
In section~\ref{s:further}, we discuss some variations and extensions of the main results as well as a related interesting issue of ``clockwork demon.''

\section{Error-free feedback}
\label{s:noerror}
Let us start from an ideal setting with no errors, and present our main result.

\paragraph*{Setting}
We consider an isolated quantum mechanical system\footnote{
The system can be a small (and literally isolated) one, or a combination of a (small) system of interest and a larger system which plays the role of the heat bath.
} whose Hamiltonian has parameters which can be controlled by an outside agent.
The system is initially prepared in the equilibrium state.
In the first stage of the time-evolution, the Hamiltonian is changed according to a fixed protocol.
Then one makes a measurement on the system.
The second stage of the time evolution takes into account feedback from the measurement, and the Hamiltonian is changed according to a protocol which depends on the outcome of the measurement.

Let us be more precise.
By $H^{(0)}$ we denote the initial Hamiltonian of the system.
Its normalized eigenstates and the corresponding eigenvalues are denoted as $\varphi_i$ and $E^{(0)}_i$, respectively, with $i=1,2,\ldots$.
The projection onto $\varphi_i$ is denoted as $P^{(0)}_i$.
We write the corresponding canonical density matrix as\footnote{
By $A:=B$ or $B=:A$, we mean that $A$ is defined in terms of $B$.
}
\eq
\rhocan^{(0)}:=\frac{e^{-\beta H^{(0)}}}{Z_0}=\sum_i\frac{e^{-\beta E^{(0)}_i}}{Z_0}P^{(0)}_i,
\lb{can0}
\en
with $Z_0:=\sum_ie^{-\beta E^{(0)}_i}$.
Finally $F^{(0)}:=-\beta^{-1}\log Z_0$ is the corresponding free energy.

Since we perform feedback, the Hamiltonian $H^{(j)}$ at the final moment depends on the outcome $j=1,\ldots,n$ of the intermediate measurement.
We denote by $\psi^{(j)}_k$ and $E^{(j)}_k$ the normalized  eigenstate and the corresponding eigenvalue, respectively, of $H^{(j)}$ with $k=1,2,\ldots$.
The projection onto $\psi^{(j)}_k$ is denoted as $P^{(j)}_k$.
Again we write the corresponding canonical density matrix as
\eq
\rhocan^{(j)}:=\frac{e^{-\beta H^{(j)}}}{Z_j}=\sum_k\frac{e^{-\beta E^{(j)}_k}}{Z_j}P^{(j)}_k,
\lb{rhoj}
\en
with $Z_j:=\sum_ke^{-\beta E^{(j)}_k}$, and the free energy as $F^{(j)}:=-\beta^{-1}\log Z_j$.

We consider the following process in the line of \cite{Hal} (see also \cite{Jorge}).
Initially the system is in the equilibrium state $\rhocan^{(0)}$.
At the initial moment, one makes a projective measurement of the energy $H^{(0)}$, whose outcome is denoted as $E_i$ or simply\footnote{
For simplicity we assume that $H^{(0)}$ has no degeneracy so that an accurate measurement of the energy uniquely determines $i$.
But this assumption is not essential.
} $i$.
Then the Hamiltonian is changed according to a fixed protocol for a certain amount of time, and the state of the system evolves by an unitary operator $U$.
This is the first stage.
Then one makes a projective measurement\footnote{
It is trivial to extend the results to the case where one makes measurement and feedback repeatedly as in \cite{JS}.
} with outcomes $j=1,\ldots,n$.
We assume that the measurement is described by a set of projection operators $\meas_1,\ldots,\meas_n$ such that $\sum_{j=1}^n\meas_j=1$ and $\meas_j\meas_{j'}=0$ if $j\ne j'$.
The rest of the time evolution, which is the second stage, depends on the outcome $j$.
The Hamiltonian is changed according to a fixed protocol associated with $j$, and the state evolves by an unitary operator $U_j$.
Finally one makes a projective measurement of the final Hamiltonian $H^{(j)}$, whose outcome is denoted as $E^{(j)}_k$ or $k$.

The probability that one gets successive outcomes $i$, $j$, $k$ in the above process is given by
\eq
\lb{pijk}
p(i\to j\to k):=\Tr[\,P^{(j)}_k\, U_j\, \meas_j\, U\, P^{(0)}_i\, U^\dagger\, \meas_j\, U_j^\dagger\,P^{(j)}_k\,]\,\frac{e^{-\beta E^{(0)}_i}}{Z_0}.
\en
One can easily verify that it is normalized as $\sum_{i,j,k}p(i\to j\to k)=1$.
We define the average with respect to $p(i\to j\to k)$ as
\eq
\lb{avdef}
\sbkt{f(i,j,k)}_p:=\sum_{i,j,k}f(i,j,k)\,p(i\to j\to k),
\en
where $f(i,j,k)$ is an arbitrary function of $i$, $j$, and $k$.

\paragraph*{Main result}
In this setting we show that 
\eq
\Bigl\langle\exp\Bigl[
\beta\bigl\{W_{i,j,k}-(F^{(0)}-F^{(j)})\bigr\}
\Bigr]\Bigr\rangle_p
=\gamma,
\lb{main}
\en
where $W_{i,j,k}:=E^{(0)}_i-E^{(j)}_k$, and
\eq
\gamma:=\sum_j\Tr[\,\meas_j\, U_j^\dagger\,\rhocan^{(j)}\, U_j\, \meas_j\,].
\lb{gamma}
\en
This is the quantum extension of one of the generalized Jarzynski relations derived by Sagawa and Ueda \cite{SU}.
The equality \rlb{main} corresponds to equation (6) in \cite{SU}.

Since $W_{i,j,k}$ is the difference between the initial and the final energy of the system, it is natural to identify it with the work done by the system.
But let us remark that this definition relies on the rather artificial setting where one precisely measures the energy in the initial and the final moments (but see the beginning of section~\ref{s:further}, where we discuss the corresponding inequality).
We also note that, in quantum systems, exchange of work also takes place  during measurement processes.

The quantity $\Tr[\,\meas_j\, U_j^\dagger\,\rhocan^{(j)}\, U_j\, \meas_j\,]$ which appears in \rlb{gamma} can be interpreted as the probability that one observes $\meas_j$ when the system starts from the equilibrium state $\rhocan^{(j)}=e^{-\beta H^{(j)}}/Z_j$ and evolves by the inverse time-evolution $U_j^\dagger$.
This probability is expected to be close to one if the time-evolution $U_j$ is chosen in such a way that any state within the range of $\meas_j$ evolves into a state not too far from the equilibrium state $\rhocan^{(j)}$.
In such a case, $\gamma$ becomes much larger than unity, suggesting that the feedback is designed in an efficient manner.
Note that $\gamma$ can be less than unity for a badly designed feedback.
See the example below.
As is stressed by Sagawa and Ueda \cite{SU}, a remarkable feature of the quantity $\gamma$ is that it can be measured by independent experiments.


\paragraph*{Derivation}
By noting that $\sum_iP^{(0)}_i=1$, and using the property of the trace, one gets
\eqa
\Bbkt{e^{\beta(E^{(0)}_i-E^{(j)}_k)}\,\frac{Z_0}{Z_j}}_p&=\sum_{i,j,k}
\Tr[\,P^{(j)}_k\, U_j\, \meas_j\, U\, P^{(0)}_i\, U^\dagger\, \meas_j\, U_j^\dagger\,P^{(j)}_k\,]\,
\frac{e^{-\beta E^{(j)}_k}}{Z_j}
\nl
&=\sum_{j,k}
\Tr[\,P^{(j)}_k\, U_j\, \meas_j\, U_j^\dagger\,P^{(j)}_k\,]\,
\frac{e^{-\beta E^{(j)}_k}}{Z_j}
\nl
&=\sum_{j,k}
\Tr[\,\meas_j\, U_j^\dagger\,P^{(j)}_k\, U_j\, \meas_j\,]\,
\frac{e^{-\beta E^{(j)}_k}}{Z_j}.
\lb{der}
\ena
By recalling \rlb{rhoj}, we see that the right-hand side is equal to $\gamma$ of \rlb{gamma}.
By noting that $Z_0/Z_j=\exp[-\beta(F^{(0)}-F^{(j)})]$, we get the desired relation \rlb{main}.

\paragraph*{Example}
As a simple illustrative example, consider a two-level system.
We set $U=1$ and $\meas_j=P^{(0)}_j$, i.e., the intermediate measurement is the same as the initial measurement of the energy.
As for the feedback, we set $U_1=1$ and let $U_2$ be the operator which simply switches the two eigenstates of $H^{(0)}$.
We also set $H^{(1)}=H^{(2)}=H^{(0)}$.
Then we have $p(1\to1\to1)=e^{-\beta E^{(0)}_1}/Z_0$ and $p(2\to2\to1)=e^{-\beta E^{(0)}_2}/Z_0$ with $Z_0=e^{-\beta E^{(0)}_1}+e^{-\beta E^{(0)}_2}$, and $p(i\to j\to k)=0$ for all other combinations.
We then find
\eq
\Bbkt{e^{\beta(E^{(0)}_i-E^{(j)}_k)}}_p=
\frac{2\, e^{-\beta E^{(0)}_1}}{e^{-\beta E^{(0)}_1}+e^{-\beta E^{(0)}_2}}.
\en

When $E^{(0)}_1<E^{(0)}_2$, the right-hand side, which is $\gamma$, is clearly larger than unity.
This reflects the fact that the demon has made a clever use of the information to get extra work from the system.
The case $E^{(0)}_1>E^{(0)}_2$, where $\gamma$ become less than unity, provides an example of a failed demon who made a wrong use of the information to lose work.

\section{Feedback with classical errors}
\label{s:error}
Next we consider a feedback process which includes errors, and extend the other relation obtained by Sagawa and Ueda.

\paragraph*{Setting}
We consider almost the same process as in section~\ref{s:noerror}, but assume that the intermediate measurement now involves errors.
We assume that the errors are of purely classical nature\footnote{
We admit that this setting is artificial.
The main motivation for studying it is that we can derive the second Jarzynski-Sagawa-Ueda relation \rlb{main3} only in this setting.
}, i.e., when the intermediate measurement (described by $\meas_1,\ldots,\meas_n$) yields the result $j$, one mis-interprets the result as $j'$ with a given probability\footnote{
In the notation of \cite{SU}, $\err(j\to j')$ should read $P[j'|j]$.
} $\err(j\to j')\ge0$.
The probability is normalized as $\sum_{j'}\err(j\to j')=1$ for any $j$.
The error-free process considered in section~\ref{s:noerror} is recovered by setting $\err(j\to j')=\delta_{j,j'}$.
In what follows we assume that for each $j'$, the probability $\err(j\to j')$ is either vanishing for all $j$ or nonvanishing for all $j$.

The rest of the process (i.e., the second stage of the time evolution and the final measurement) is executed according to the result $j'$ (not $j$).
This means that the probability \rlb{pijk} is modified as
\eq
\tp(i\to j\to j'\to k):=
\Tr[\,P^{(j')}_k\, U_{j'}\, \meas_j\, U\, P^{(0)}_i\, 
U^\dagger\, \meas_j\, U_{j'}^\dagger\,P^{(j')}_k\,]\,
\err(j\to j')\,\frac{e^{-\beta E^{(0)}_i}}{Z_0}.
\lb{pijjk}
\en
As in \rlb{avdef}, we denote the average over this probability as
\eq
\sbkt{f(i,j,j',k)}_{\tp}:=\sum_{i,j,j',k}f(i,j,j',k)\,\tp(i\to j\to j'\to k).
\en
Let us also define
\eq
\tp_2(j):=\sum_{i,j',k}\tp(i\to j\to j'\to k),\quad
\tp_3(j'):=\sum_{i,j,k}\tp(i\to j\to j'\to k),
\en
and
\eq
\tp_{2,3}(j,j'):=\sum_{i,k}\tp(i\to j\to j'\to k),
\en
which are the probabilities to observe $j$, to observe $j'$, and to observe a pair $(j,j')$, respectively.
As in \cite{SU}, we define the (unaverage) mutual information as
\eq
I_{j,j'}:=\log\frac{\err(j\to j')}{\tp_3(j')}
=\log\frac{\tp_{2,3}(j,j')}{\tp_2(j)\,p_3(j')},
\lb{Ijj}
\en
where the second equality comes from $\tp_{2,3}(j,j')=\tp_2(j)\,\err(j\to j')$.
By averaging this, one gets
\eq
\sbkt{I_{j,j'}}_{\tp}=
\sum_{j,j'}\tp_{2,3}(j,j')\,\log\frac{\tp_{2,3}(j,j')}{\tp_2(j)\,\tp_3(j')},
\lb{I}
\en
which is the mutual information.

\paragraph*{Main results}
In this setting we show that
\eq
\Bigl\langle\exp\Bigl[
\beta\bigl\{
W_{i,j',k}-(F^{(0)}-F^{(j')})\bigr\}-I_{j,j'}
\Bigr]\Bigr\rangle_{\tp}=1,
\lb{main3}
\en
where the ``work'' is defined by  $W_{i,j',k}:=E^{(0)}_i-E^{(j')}_k$ as before.
This is a quantum version of the other generalized Jarzynski relation, equation (4) in \cite{SU}, derived by Sagawa and Ueda.
Interestingly, this type of equality seems to be derivable only in the present setting with classical errors.
Indeed the mutual information \rlb{Ijj}, \rlb{I} is a purely classical quantity as opposed to the QC-mutual information (see \cite{SU1}), which takes into account the full quantum nature of the system\footnote{
We note, however, that our equality \rlb{main3} is valid as it is when we take into account quantum mechanical errors by measurement operators $M_j$ which satisfy a certain condition.
See section~\ref{s:further}.
}.
See the discussion about the error-free limit at the end of the present section, and about the corresponding inequality in section~\ref{s:further}.

In the present setting with errors, we can also derive a relation which is exactly the same as \rlb{main} but $\gamma$ replaced by
\eq
\tgamma:=\sum_{j,j'}\err(j\to j')\,\Tr[\,\meas_j\, U_{j'}^\dagger\,\rhocan^{(j')}\, U_{j'}\, \meas_j\,].
\lb{gamma2}
\en

\paragraph*{Derivation}
As in \rlb{der}, we use $\sum_iP^{(0)}_i=1$ and $\sum_j\meas_j=1$ to observe that
\eqa
\Bbkt{e^{\beta(E^{(0)}_i-E^{(j')}_k)}\,\frac{Z_0}{Z_{j'}}\,
\frac{\tp_3(j')}{\err(j\to j')}}_{\tp}&
=\sum_{i,j,j',k}
\Tr[\,P^{(j')}_k\, U_{j'}\, \meas_j\, U\, P^{(0)}_i\, U^\dagger\, \meas_j\, U_{j'}^\dagger\,P^{(j')}_k\,]\,
\frac{e^{-\beta E^{(j')}_k}}{Z_{j'}}\,\tp_3(j')
\nl
&=\sum_{j,j',k}
\Tr[\,P^{(j')}_k\, U_{j'}\, \meas_j\, U_{j'}^\dagger\,P^{(j')}_k\,]\,
\frac{e^{-\beta E^{(j')}_k}}{Z_{j'}}\,\tp_3(j')
\nl
&=\sum_{j',k}
\Tr[\,P^{(j')}_k\,]\,
\frac{e^{-\beta E^{(j')}_k}}{Z_{j'}}\,\tp_3(j')
=\sum_{j'}\tp_3(j')=1,
\lb{der2}
\ena
where we noted $\Tr[\,P^{(j')}_k\,]=1$.
By recalling the definition \rlb{Ijj}, we get \rlb{main3}.

The derivation of \rlb{main} with $\gamma$ replaced by \rlb{gamma2} is essentially the same as that in section~\ref{s:noerror}, and is omitted.

\paragraph*{Error-free limit}
Consider the limit where $\err(j\to j')$ tends to $\delta_{j,j'}$, where one recovers the error-free setting of section~\ref{s:noerror}.
The (unaverage) mutual information \rlb{Ijj} becomes $I_{j,j'}\to \delta_{j,j'}\,S_j$ with $S_j=-\log\tp_2(j)=-\log\tp_3(j)$.
Thus the averaged quantity \rlb{I} becomes $\sbkt{I_{j,j'}}_{\tp}\to\sbkt{S_j}_p=-\sum_j\tp_2(j)\log\tp_2(j)$, which is the Shannon entropy.

Then one might be tempted to conjecture the validity of a relation corresponding to \rlb{main3} for the error-free setting of section~\ref{s:noerror}, namely,
\eq
\Bigl\langle\exp\Bigl[
\beta\bigl\{W_{i,j,k}-(F^{(0)}-F^{(j)})\bigr\}-S_{j}\Bigr
]\Bigr\rangle_p
\stackrel{?}{=}1.
\lb{wrong}
\en
Unfortunately, a careful look at the above derivation  reveals that this is not valid in general\footnote{
It is a tacit assumption in \cite{SU} that $P[y|\Gamma_m]$ is always nonvanishing.
Although the example of the Szilard engine considered in \cite{SU} fails to satisfy the assumption, the relation (4) happens to be valid for certain reasons specific to the model.
We thank Takahiro Sagawa for clarifying this point.
}.
In \rlb{der2}, we are making use of the relation $\err(j\to j')/\err(j\to j')=1$, which is not true in the error-free limit.
In fact one can easily check that the conjectured \rlb{wrong} does not hold in the two-level system considered at the end of section~\ref{s:noerror}.

\section{Further observations and discussions}
\label{s:further}
\paragraph*{Corresponding inequalities}
As is always the case, one can use Jensen's inequality $\sbkt{e^f}\ge e^{\sbkt{f}}$ to derive inequalities from the equalities that we have shown.
From \rlb{main}, in particular, we get
\eq
\sbkt{W_{i,j,k}}_p
\le F^{(0)}-\sum_jp_2(j)\,F^{(j)}+\frac{1}{\beta}\log\gamma,
\lb{ineq}
\en
where $p_2(j):=\sum_{i,k}p(i\to j\to k)=\Tr[\, \meas_j\, U\, \rhocan^{(0)}\, U^\dagger\, \meas_j\,]$ is the probability that one gets an outcome $j$ in the intermediate measurement.
Note that the left-hand side is written as
\eq
\sbkt{W_{i,j,k}}_p=
\Tr[H^{(0)}\rhocan^{(0)}]-\sum_jp_2(j)\Tr[H^{(j)}\rho_\mathrm{fin}^{(j)}],
\lb{Wp}
\en
where
\eq
\rho_\mathrm{fin}^{(j)}:=
\frac{U_j\meas_j\,U\,\rhocan^{(0)}\,U^\dagger\,\meas_j\,U_j^\dagger}
{\Tr[\, \meas_j\, U\, \rhocan^{(0)}\, U^\dagger\, \meas_j\,]}
\en
is the final state of the system when the outcome of the intermediate measurement is $j$.
It should be remarked that \rlb{Wp} only involves standard quantum mechanical expectation values in the initial and the final states.
This is in contrast with the left-hand side of the  equality \rlb{main}, which is defined only  in  a rather artificial setting where one measures the energy both in the initial and the final moments.
Since  \rlb{Wp} is the expectation value of the total work done by the system\footnote{
As we have noted before this includes work exchanged during the measurement.
}, the inequality \rlb{ineq} can be interpreted as a generalization of the second law of thermodynamics.
This is distinct from the second law derived by Sagawa and Ueda in \cite{SU1}, which is valid for quantum systems with feedback.

Obliviously one can replace $e^{\beta(E^{(0)}_i-E^{(j)}_k)}$ in the left-hand side of \rlb{der} by $e^{\beta E^{(0)}_i-\beta_k E^{(j)}_k}$ with an arbitrary $\beta_1,\ldots,\beta_n$ to get an exact relation.
This relation again yields an inequality for $\Tr[H^{(0)}\rhocan^{(0)}]$ and $\Tr[H^{(j)}\rho_\mathrm{fin}^{(j)}]$, which is more general than \rlb{ineq}.
It would be interesting to see whether one can get stronger inequalities than  \rlb{ineq} by choosing optimal $\beta_j$ which reflect the nature of the operation and the feedback.

From the equality \rlb{main3}, one gets another second law
\eq
\sbkt{W_{i,j',k}}_{\tp}
\le F^{(0)}-\sum_{j'}\tp_3(j')\,F^{(j')}+\frac{1}{\beta}\sbkt{I_{j,j'}}_{\tp},
\lb{ineq2}
\en
where 
\eq
\sbkt{W_{i,j',k}}_{\tp}=\Tr[H^{(0)}\rhocan^{(0)}]-
\sum_{j,j'}\tp_{2,3}(j,j')\Tr[H^{(j')}
\tilde{\rho}_\mathrm{fin}^{(j,j')}]
\en
is again the expectation value of the total work, where we wrote
\eq
\tilde{\rho}_\mathrm{fin}^{(j,j')}:=
\frac{U_{j'}\meas_j\,U\,\rhocan^{(0)}\,U^\dagger\,\meas_j\,U_{j'}^\dagger}
{\Tr[\, \meas_j\, U\, \rhocan^{(0)}\, U^\dagger\, \meas_j\,]}.
\en
The inequality \rlb{ineq2} looks identical to the second law for quantum systems with feedback derived by Sagawa and Ueda \cite{SU1}, but their inequality contains the QC-mutual information rather than the classical mutual information $\sbkt{I_{j,j'}}_{\tp}$.
The two inequalities are indeed different relations\footnote{
If one applies the result of \cite{SU1} to our setting, one gets an inequality which takes into account ``quantum mechanical errors'' but is independent of the classical error probability $\err(j\to j')$ \cite{S}.
On the other hand our inequality (which,  as we explain below, is valid as it is if the measurement is described by operators $M_j$ which satisfy a certain condition) is apparently insensitive to quantum mechanical errors. 
}.
We also note that one can safely take the error-free limit in the inequality \rlb{ineq2} since nothing like $0/0$ is encountered here.

\paragraph*{General measurements}
Let us remark that the same derivation as in section~\ref{s:noerror} works if one replaces the projections $\meas_1,\ldots,\meas_n$ with general measurement operators $M_1,\ldots,M_n$ which satisfy $\sum_jM_j^\dagger M_j=1$.
One gets the Jarzynski-Sagawa-Ueda relation \rlb{main} and the corresponding inequality \rlb{ineq} with $\gamma$ replaced by
\eq
\gamma:=
\sum_j\Tr[\,M_j^\dagger\, U_j^\dagger\,\rhocan^{(j)}\, U_j\, M_j\,].
\lb{gammaM}
\en
Note that the summand is not the probability of observing $M_j$ in the state $ U_j^\dagger\,\rhocan^{(j)}\, U_j$ unless $M_j$ is Hermitian, since in \rlb{gammaM}  we have $M_j^\dagger$ instead of $M_j$.
It is expected that $M_j^\dagger$ corresponds to a suitable ``time-reversed'' measurement.
See, for example, \cite{TU}.

The relation \rlb{main3}, on the other hand, can be derived (as it is) only when one has $\sum_jM_jM_j^\dagger=1$.
This condition is valid when all $M_j$ are Hermitian, or (more generally) when $[M_j,M_j^\dagger]=0$ for all $j$, but is not valid in general.

\paragraph*{Quantum Jarzynski relation with measurement}
Consider the error-free setting of section~\ref{s:noerror}, and further suppose that the agent makes a measurement, but do not perform any feedback.
This means that we take $U_j=U'$, $H^{(j)}=H'$, and $\rhocan^{(j)}=\rhocan'$ for any outcome $j=1,\ldots,n$.

In this case we see from \rlb{gamma} that
\eq
\gamma=\sum_j\Tr[\,\meas_j\, (U')^\dagger\,\rhocan'\, U'\, \meas_j\,]=\Tr[\,(U')^\dagger\,\rhocan'\, U'\,]=\Tr[\,\rhocan'\, ]=1,
\en
and hence our main result \rlb{main} reduces to the quantum version of Jarzynski relation \cite{Hal}.
Since a measurement process generally disturbs a quantum state, it is a nontrivial fact that the quantum Jarzynski relation gets no modifications here.
This is closely related to the extension of fluctuation theorem in \cite{CTH}.

When projective measurement is replaced by more general measurement as above, we have $\gamma=1$ only when the condition $\sum_jM_jM_j^\dagger=1$ is valid.

If one takes the setting of section~\ref{s:error} with errors, and assumes that $\err(j\to j')$ is independent of $j$, one immediately gets $\tgamma=1$ from \rlb{gamma2}.
Thus we also have the standard Jarzynski relation in this case.
This is natural since one has $p_3(j')=\err(j\to j')$ and hence $I_{j,j'}=0$, i.e., the agent is making no use of information.

\paragraph*{On ``clockwork demon''}
We have here assumed the existence of an external intelligent agent (demon) who performs feedback taking into account the outcome of the intermediate measurement.
It is also interesting to consider the possibility of  designing an external quantum mechanical system, a ``clockwork demon'',  which is programmed to perform the exact feedback we want\footnote{
The following discussion of course applies to the classical setting (as in \cite{SU}) as well.
We also stress that this is quite a standard thought, which has appeared in many different contexts.
}.

The Hamiltonian of the whole system is written as $H_\mathrm{tot}(t)=H_\mathrm{system}(t)+H_\mathrm{demon}(t)+H_\mathrm{int}(t)$, and its time-dependence is fixed in advance.
The interaction Hamiltonian $H_\mathrm{int}(t)$ is vanishing in the initial and the final moments.
Initially both the system and the demon are in their (canonical) equilibrium states\footnote{
In many cases, it is more natural to assume that the demon is initially in a quasi-equilibrium state.
More precisely, one assumes that the state of demon is initially restricted to a certain subspace of the whole Hilbert space, and distributed according to the canonical distribution within the subspace.
Then the original Jarzynski relation does not apply, and the following discussion is not valid as it is.
} with a common $\beta$.
Then the demon Hamiltonian $H_\mathrm{demon}(t)$ and the interaction Hamiltonian $H_\mathrm{int}(t)$ are changed to let the demon interact with the system as designed.
The actual design of such an autonomous feedback system is far from trivial.
It is indeed a nontrivial question whether one can realize an arbitrary feedback scenario in this manner, but we won't go into details here.

It is crucial that the original Jarzynski relation without feedback \cite{Chris,Hal} applies to this situation.
Since the system and the demon are decoupled in the initial and the final moments, the relation reads
\eq
\Bigl\langle\exp\Bigl[\beta\bigl\{W-(F_\mathrm{system}^\mathrm{init}-F_\mathrm{system}^\mathrm{fin})\bigr\}\Bigr]\Bigr\rangle=\exp\bigl[\beta(F_\mathrm{demon}^\mathrm{init}-F_\mathrm{demon}^\mathrm{fin})\bigr],
\lb{main2}
\en
where $W$ is the difference between the initial total energy and the final total energy.
We here make a nontrivial and crucial assumption that the clockwork demon is designed in such a way that it does not exchange appreciable work with the system or with the agent who changes the Hamiltonian\footnote{
We still do not know in which case this (rather tricky) assumption is realized.
}.
Then the work $W$ is interpreted as essentially the work done by the system to the external agent, and the left-hand side of \rlb{main2} can be identified with that of \rlb{main}.
Comparing \rlb{main2} with \rlb{main}, one sees that $\gamma$, in this case, is directly related to the free energy difference of the clockwork demon.

Let us stress, however, that this consideration does not diminish the significance of Sagawa and Ueda's generalization nor reduce it to the original Jarzynski relation.
The power of the generalized relation appears, for example, in the compact representation \rlb{gamma} of $\gamma$, which does not only suggest a clear interpretation but also makes it possible to measure $\gamma$ experimentally.
One can say that, by considering an idealized setting with feedback, Sagawa and Ueda were able to sort, in quite a neat manner, complicated exchange of heat and work into the part which is directly related to the feedback and the part intrinsic to the system.

\bigskip

{\small 
We wish to thank Takahiro Sagawa for indispensable discussions and comments, Jordan Horowitz and Akira Shimizu for useful discussions.
}



\begin{thebibliography}{10}

\bibitem{ECM}
D. J. Evans, E. G. D. Cohen, and G. P. Morriss,
Probability of second law violations in shearing steady states,
Phys. Rev. Lett. {\bf 71}, 2401--2404 (1993).

\bibitem{Giovanni}
G. Gallavotti and E. G. D. Cohen,
Dynamical Ensembles in Nonequilibrium Statistical Mechanics,
Phys. Rev. Lett. {\bf 74}, 2694--2697 (1995), {\tt chao-dyn/9410007}.


\bibitem{Crooks}
G. E. Crooks, 
Entropy production fluctuation theorem and the nonequilibrium work relation for free energy differences,
Phys. Rev. {\bf E60}, 2721--2726 (1999),
{\tt  cond-mat/9901352}.


\bibitem{Chris}
C. Jarzynski,
Nonequilibrium Equality for Free Energy Differences,
Phys. Rev. Lett. {\bf 78}, 2690 (1997), {\tt cond-mat/9610209}.

\bibitem{SU1}
T. Sagawa and M. Ueda,
Second Law of Thermodynamics with Discrete Quantum Feedback Control,
Phys. Rev. Lett. {\bf 100}, 80403 (2008), {\tt  arXiv:0710.0956}.

\bibitem{SU}
T. Sagawa and M. Ueda,
Generalized Jarzynski Equality under Nonequilibrium Feedback Control,
Phys. Rev. Lett. {\bf 104}, 90602 (2010), {\tt arXiv:0907.4914}.

\bibitem{Toyabe}
S. Toyabe, T. Sagawa, M. Ueda, E. Muneyuki, and M. Sano,
Experimental demonstration of information-to-energy conservation and validation of the generalized Jarzynski equality,
Nature Phys. {\bf 6}, 988--992 (2010), {\tt arXiv:1009.5287}.

\bibitem{CTH}
M. Campisi, P. Talkner, P. H{\"a}nggi,
Fluctuation Theorems for Continuously Monitored Quantum Fluxes,
Phys. Rev. Lett. {\bf 105}, 140601 (2010),
{\tt arXiv:1006.1542}.

\bibitem{JS}
J. M. Horowitz and S. Vaikuntanathan,
Nonequilibrium Detailed Fluctuation Theorem for Repeated Discrete Feedback,
preprint (2010), {\tt arXiv:1011.4273}.

 
\bibitem{Hal}
H. Tasaki,
Jarzynski relations for quantum systems and some applications,
unpublished note (2000), {\tt cond-mat/0009244}.

\bibitem{Jorge}
J. Kurchan, A quantum fluctuation theorem,
preprint (2000), {\tt cond-mat/0007360}.

\bibitem{S}
T. Sagawa, private communication.

\bibitem{TU}
H. Terashima and M. Ueda,
Hermitian conjugate measurement,
Phys. Rev. {\bf A81}, 1094--1622 (2010), 
{\tt arXiv:0709.1210}.

\end{thebibliography}
\end{document}